\documentclass[twocolumn,showpacs,preprintnumbers,amsmath,amssymb]{revtex4}
\usepackage{graphicx,epsf}
\begin{document}

\title{Jamming coverage in competitive random sequential adsorption of a binary mixture}
\author{M. Kamrul Hassan$^{1,2}$, J\"urgen Schmidt$^1$, Bernd Blasius$^1$, and J\"urgen Kurths$^1$}
\affiliation{
$^1$University of Potsdam, Department of Physics, Postfach 601553, D-14415 Potsdam, Germany\\
$^2$ University of Dhaka, Department of Physics, Theoretical Physics Division, Dhaka 1000, Bangladesh
}
\email{khassan@agnld.uni-potsdam.de}
\date{\today}

\begin{abstract}%
We propose a generalized car parking problem where cars of two different sizes  are sequentially 
parked on a line with a given probability $q$. The free parameter $q$ interpolates between the classical 
car parking problem of only one car size and the competitive random sequential adsorption (CRSA) of a
 binary mixture. We give an exact solution to the CRSA rate equations and find that the final coverage, 
the jamming limit, of the line is always larger for a binary mixture than for the uni-sized case. 
The analytical results are in good agreement with our direct numerical simulations of the problem.
\end{abstract}


 \maketitle

The adsorption of particles onto a solid substrate is a common phenomenon and it is of wide interest in 
physics, chemistry, biology,  and in other branches of science and technology. Examples include adsorption of 
gas molecules, colloidal particles, polymer chains, bacteria, proteins, and latex particles 
\cite{kn.has1,kn.has2,kn.has3}. Due to its wide range of applications, adsorption has been studied 
experimentally, numerically, and analytically (see for example \cite{kn.has11} for a review). 

The simplest model one can think of, still capturing the generic features of the adsorption phenomenon, 
is the kinetics of {\it random sequential adsorption} (RSA) of particles of a fixed size on a d-dimensional
 substrate. The one-dimensional continuum version of the RSA process is popularly known as the 
random car parking (RCP) problem \cite{kn.has4}. 
In comparison to monolayer growth by RSA of monodisperse particles, very little attention has been given to the
similar monolayer growth by two or more species of different size, despite the fact that the latter problem is 
much closer to the real life situation than the former. Nonetheless, there has been an increasing
interest in the study of RSA of mixtures of different degree 
\cite{kn.meakin,kn.has5,kn.has6,kn.satorras,kn.has7,kn.bonnier}.
It is worth mentioning that the problem is simplified
in two extreme cases, namely, the RSA of uni-sized particles and that of mixture of particles of all sizes
obeying a power-law size distribution \cite{kn.has7}.
In the former case the ultimate structure in the long time limit is  described by the jamming
coverage, whereas in the latter case the resulting monolayer is uniquely quantified by the fractal dimension $D_f$.
The complexity arises in between these two extreme cases when the mixture contains more than one species.
The CRSA of a binary mixture has recently received new interest in the literature, 
where the properties of the jamming coverage 
obtained in analytical models \cite{kn.bonnier} and in computer simulations \cite{kn.meakin} have been discussed 
controversially.


In this rapid communication, we consider the problem of the CRSA of a binary mixture on a one-dimensional substrate, 
i.e.\ the generalization of the classical RCP problem to two car species. To this end we develop an 
analytical model in terms of rate equations for the gap size distribution function and compute expressions for 
the jamming limit of both particle species, for a size ratio $\leq 2$. We compare our results to direct 
numerical simulations of the CRSA process. 

One time step of the CRSA process for a binary mixture is defined as:
\begin{itemize}
\item[{\it i)}] A particle is selected to be deposited on the substrate: With probability $q$ the size of the particle is 
$\sigma$ and with probability $p=1-q$ its size is $m \sigma$, where $m>1$.
\item[{\it ii)}] Randomly a position is chosen on the substrate.
\item[{\it iii)}] The particle is deposited on the substrate if this position is not occupied by another particle and if 
there is no overlap with particles left and right to the position.  
\item[{\it iv)}] The process is repeated until no more gaps with size larger than $\sigma$
are left on the substrate.
\end{itemize}
Our main results are as follows. The jamming coverage of the binary 
mixture always exceeds its monodisperse counterpart. The dynamics close to the jamming limit is governed by the 
dynamics of the smaller species. The total number density, as well as the number density of the smaller particles, 
gain their asymptotic values algebraically, following Feder's law. The large particles reach their asymptotic coverage 
exponentially, with a decay constant $\sim (m-1)q$, multiplied by an algebraic prefactor $t^{-1}$.

We define the gap size distribution function 
\begin{equation}
  \label{eq:gab_p}
  P(x,t)=\frac{\sigma}{L}N(x,t)
\end{equation}
where $N(x,t)$ is the total number of gaps with size $x$ (say, in an interval $x+dx$) at time $t$ on a substrate with
 length $L$. We scale length and time as
\begin{equation}
  \label{eq:tlscale}
  x\rightarrow\frac{x}{\sigma},\quad t\rightarrow t\frac{\alpha}{L/\sigma}\,.
\end{equation}
where $\alpha$ is the rate of particles brought to the substrate. In the limit $L\rightarrow\infty$, keeping the particle 
inflow rate per unit length $\alpha/L$ fixed, the time evolution of the gap size distribution function is then 
described by the dimensionless rate equations
\begin{eqnarray}  
\label{eq:gap_1}
 {{\partial P(x,t)}\over{\partial t}} & = & -(x-\{m p +q\})P(x,t) \\
 & & + 2q\int_{x+1}^\infty P(y,t) dy 
+2p\int_{x+m}^\infty P(y,t)dy,\nonumber\\
&&\mbox{for} \quad m<x<\infty\nonumber \\
\label{eq:gap_2}
{{\partial P(x,t)}\over{\partial t}} & = & -q\left(x-1\right)P(x,t)  \\ &+ & 
2q \int_{x+1}^\infty  P(y,t)dy+2p\int_{x+m}^\infty P(y,t)dy, \nonumber \\
&&\mbox{for}\quad 1<x\le m\nonumber \\
\label{eq:gap_3}
{{\partial P(x,t)}\over{\partial t}} & =& 2q \int_{x+1}^\infty  P(y,t)dy+2p\int_{x+m}^\infty P(y,t)dy, \\
&&\mbox{for} \quad 0<x\le1\,.\nonumber
\end{eqnarray}
These equations describe a RSA process with two particle species competing for adsorption. The strength of 
competition is determined by $q$. The free parameters of the model are $q$ and $m$.

To solve the kinetic Eqs. (\ref{eq:gap_1}-\ref{eq:gap_3}), we seek a solution in the domain $x>m$ of the form 
\begin{equation}
  \label{eq:trial}
 P(x,t) = A(t) e^{-(x-\bar\sigma)t}\,,
\end{equation}
where the average size of incoming particles is expressed as 
\begin{equation}
  \label{eq:siagav}
\bar\sigma\equiv m p + q
\end{equation}
and $A(t)$ is to be determined. Substituting Eq. (\ref{eq:trial}) into Eq. (\ref{eq:gap_1}) we obtain 
\begin{equation}
\label{eq:dadt}
{{d \ln A(t)}\over{dt}}=2q{{e^{-t}}\over{t}}+2p{{e^{-m t}}\over{t}} \,.
\end{equation}
At the beginning of the RSA process the number {\it per unit length} of gaps of any length $x$ is zero. Also, initially 
there is only one gap of the size of the substrate. Solving Eq. (\ref{eq:dadt}) subject to these initial conditions,
 i.e.\
\begin{equation}
 \int_0^\infty dx P(x,0) =0, \quad\lim_{t \longrightarrow 0}\int_0^\infty dx\, x P(x,t)=1\,
\end{equation}
gives 
\begin{equation}
A(t)=t^2F(t)\,,
\end{equation}
where
\begin{equation}
F(t)= e^{-2\int_0^{t}du\{q(1-e^{-u})+p(1-e^{-mu})\}/u}.
\end{equation}
Thus, the solution for the gap size distribution function for gaps $x>m$ is
\begin{equation}
\label{eq:solsiggm}
P(x,t)=t^2 F(t)e^{-(x-\bar\sigma)t},\, x>m
\end{equation}
In the case $m\le 2$ only the solution (\ref{eq:solsiggm}) for $x>m$ contributes to the integrals in Eq. 
(\ref{eq:gap_2}). Thus, in this case a solution in the domain $1< x\le m$ can be derived in the form
\begin{equation}
P(x,t)=B(x,t)e^{-q(x-1)t}.
\end{equation}
The result is ($1< x\le m$)
\begin{eqnarray}
\label{eq:solxlm}
P(x,t) & = & 2e^{-q(x-1)t}\nonumber\\
&& \int_0^t ds F(s)s(q e^{(mp-1)s}+p e^{-qms})e^{-xps}\,.
\end{eqnarray}
For $m>2$ the yet unknown solution to Eq. (\ref{eq:gap_2}) contributes to the integrals in that equation as well, 
and thus, the integration of the equation is more difficult.

From the knowledge of the gap size distribution function in the domain $x>1$ we find the total coverage $\theta_T(t)$ by 
the adsorbing particles at time $t$ 
\begin{equation}
\theta_T(t)=1-\int_0^\infty xP(x,t)dx.
\end{equation} 
For $0<q<1$ the total coverage  $\theta_T(t)=\theta_S(t)+\theta_L(t)$ has contributions by the small particles 
$\theta_S(t)$ and large particles $\theta_L(t)$. Practically, we first compute the time derivative $\dot\theta_T(t)$,
 using Eqs. (\ref{eq:gap_1}-\ref{eq:gap_3}). Then we formulate the resulting expression in terms of $P(x,t)$ 
in the domains $x> m$ (\ref{eq:solsiggm}) and $1< x\le m$ (\ref{eq:solxlm}) alone, by means of partial integrations. 
Thus, the solution to the rate equations for $x<1$ is not needed. The contribution of the small particles $\theta_S(t)$ 
to the coverage at time $t$ is
\begin{eqnarray}
\label{eq:jam_small}
\theta_S(t)& = & 2 \int_0^t ds\,\frac{F(s)}{q t + p s} e^{-s(p+m q)}\left(p e^{-(m-1)s} + q \right)\nonumber \\
&&\left(s e^{(m-1)q(s-t)}  - s (q e^{(m-1)s} +p) + q t ( e^{(m-1) s} - 1)\right)\nonumber \\
&&+q \int_0^t e^{-(m-1)qs}\{1+(m-1)s\}F(s)ds
\end{eqnarray}
and the large particles contribute
\begin{equation}
\label{eq:jam_large}
\theta_L(t)=pm\int_0^tF(s)e^{-(m-1)qs}ds\,.
\end{equation}
We calculate the jamming limit of our model by numerical integration of Eqs. (\ref{eq:jam_small},\ref{eq:jam_large}).
 For $m=1$, we recover the classical RCP result \cite{kn.has4}
\begin{equation}
  \label{eq:renyi}
  \theta_T(t)=\int_0^tF(s)ds\,,
\end{equation}
which equals $\theta_R = 0.748...$ (the Renyi-limit) for $t\rightarrow\infty$.


To test our analytic results we simulate the CRSA problem on a computer. Naturally, the simulations restrict on 
finite substrate lengths $L$. However, for sufficiently large lengths $L\gg\sigma$ the effects of finite $L$ are small. 
The algorithm we use follows literally the steps {\it i)} -- {\it iv)} described in the introduction. The time scale in 
the simulations is set by the number of particles brought to the substrate (successful or not). 
This discrete `loop index' $i$ relates to the time scale of the rate equations $t$ as
\begin{equation}
t = i \frac{\sigma}{L}\,.
\end{equation}
Using this time scale and expressing all lengths in the simulation in terms of $\sigma$, the simulational results 
can be directly compared to the solution to the rate equations and Eqs. (\ref{eq:jam_small}, \ref{eq:jam_large}) 
for the jamming limit.

\begin{figure}[!htb]
  \centerline{\includegraphics[width=8.5cm]%
 {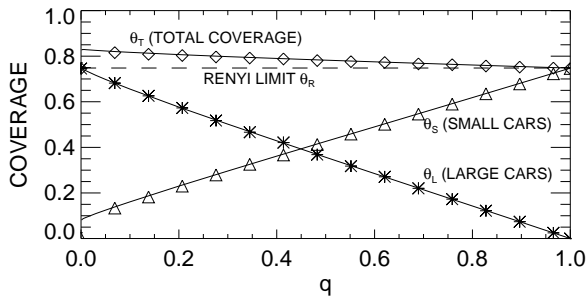}}
 \caption{Coverage in the jamming limit vs $q$ for $m = 1.5$, with the respective contributions from small and large 
particles. Symbols represent simulational results, lines the analytical solution (\ref{eq:jam_small},\ref{eq:jam_large}).
 Simulations were carried out for a substrate length of 10000 small cars, each symbol is an average over 20 independent 
realizations.}
  \label{fig:1}
\end{figure}

In Fig. (\ref{fig:1}) results for the jamming coverage are shown for the case $m = 1.5$, where the parameter $q$ is 
varied.  
In general, we find an excellent agreement between simulation and theory. The total coverage is found to be always 
larger for binary mixtures ($q \ne 0,\,q\ne 1$) than for the uni-sized case ($q = 0$ or $q = 1$). Also, the 
contribution of the small cars to the jamming limit does not vanish as $q \rightarrow 0$. This is intuitively clear,
 since for a very small $q$ first the large particles may reach their jamming limit $\theta_R=0.748$, while the small
 particles then have time to gradually fill the remaining gaps. In this sense the adsorption 
processes of large and small particles decouple in the limit $q \rightarrow 0$. 
The CRSA process is thus particularly effective in covering the substrate for small $q$ and hence 
there is a sudden jump from $\theta_T(q=0)$ to $\theta_T(q=0^+)$.
 
\begin{figure}[!htb]
\centerline{\includegraphics[width=8.5cm]%
{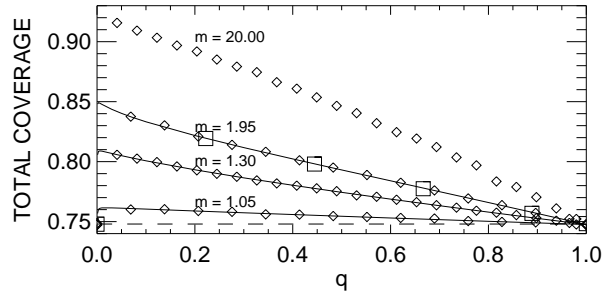}}
 \caption{The total coverage vs $q$ for various values of $m$. Diamond symbols refer to simulations with a substrate 
length of 10000 small cars, and for the case $m = 1.95$ results from a simulation with $L = $ 5000 are shown for 
comparison as square symbols.}
  \label{fig:2}
\end{figure}

The highest coverages are obtained for small $q$ and a large size ratio $m$ of the binary mixture 
(see Fig. (\ref{fig:2})). In this case the large particles rapidly cover a fraction $\theta_R$ of the substrate,
 while the small particles then fill again the same fraction of the remaining gaps. Then 
the maximum possible coverage $\theta_{max}=\theta_R(2-\theta_R)=0.937$ may be reached in the limit $m\longrightarrow$. 
The simulation with $m = 20$ 
shown in Fig. (\ref{fig:2}) is already very close to this limit. 
Our theoretical solution for the gap size distribution function restricts on the domain $m\le 2$, thus no theory curve 
is plotted for the latter simulation. As $q$ increases the coverage $\theta_T$ decreases due to increasing strength
of competition for adsorption between the two species. This can be explained as follows. 
Every time a small particle wins in the 
competition it occupies less territory than a large particle would. Moreover, every occupation
by the smaller specie creates an exclusion zone for further adsorption of larger particles. 
In the limit $q\longrightarrow 1$
once again the two processes decouple but now the remaining gaps are smaller than the large
particles and the symmetry is broken i.e. $\theta_T(q\longrightarrow 0)\neq \theta_T(q\longrightarrow 1)$.
Note, that the CRSA model has no smooth transition in the limit
 $m\rightarrow\infty$ to the competitive adsorption model of a binary mixture of point particles and finite 
sized particles; the small particles of finite length will always occupy a fraction $\theta_R$ of the gaps 
between the large particles. Thus, the total coverage $\theta_T$ is always larger than $\theta_R$. Only if 
their length is indeed zero the small particles will not contribute to the coverage, and their only effect is 
to prevent the large particles from reaching their maximum coverage $\theta_R$. In this case the total coverage 
stems from the large particles alone (see Fig. (\ref{fig:1})), and it remains always below the Renyi-limit $\theta_R$
\cite{kn.has10}.

\begin{figure}[!htb]
\centerline{\includegraphics[width=8.5cm]%
{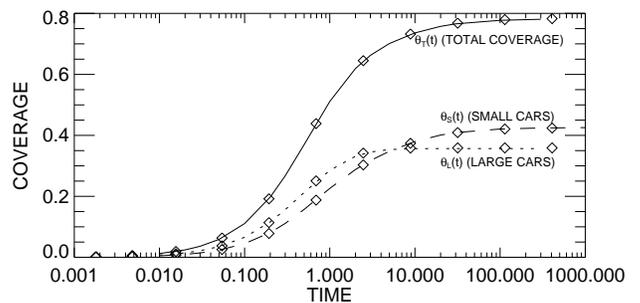}}
 \caption{Time development of the coverage for $q=0.5$, $m = 1.5$. Lines (solid, dashed, dotted) are theoretical results,
 symbols from simulations.}
  \label{fig:3}
\end{figure}
\begin{figure}[!htb]
\centerline{\includegraphics[width=8.5cm]%
{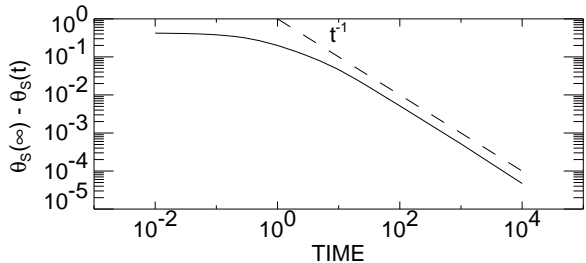}}
 \caption{The approach of $\theta_S(t)$, and thus $\theta_T(t)$, to its asymptotic value obeys Feder's law (equation 
(\ref{eq:feders_law})).}
  \label{fig:4}
\end{figure}

The time development of the coverage is shown in Fig. (\ref{fig:3}) for the case $q=0.5$, $m = 1.5$. Again a good
 agreement is found for theoretical and simulational results. The contribution of the large particles to the total 
coverage reaches its final value very rapidly. Asymptotically, Eq. (\ref{eq:jam_large}) yields
\begin{eqnarray}
  \label{eq:asym_large}
 \theta_L(\infty)-\theta_L(t) &=& pm\int_t^\infty F(s)e^{-(m-1)qs}ds\nonumber\\
&\sim&\frac{e^{-(m-1)qt}}{t},\quad \mbox{as}\,\,t\rightarrow\infty\,.
\end{eqnarray}
Thus, the asymptotic behavior of the coverage, reaching eventually the jamming limit, is dominated by the small
 particles. Indeed, the contribution of the small particles to the total coverage (Eq. (\ref{eq:jam_small}))
 approaches its final value algebraically as $t^{-1}$ (see Fig. (\ref{fig:4})).
Thus, we verified the validity of Feder's law for this system
\begin{equation}
 \label{eq:feders_law}
  \theta_d(\infty)-\theta_d(t) \sim t^{-1/d}\,,
\end{equation}
where d is the dimension of the substrate.

The present work provides an exact theoretical basis on RSA of mixtures and
removes the controversy between the numerical simulations presented by Meakin and Jullien 
\cite{kn.meakin} for the {\it two-dimensional} CRSA of disks and that of Bonnier \cite{kn.bonnier} in 1d. 
Meakin and Jullien report as well a larger 
coverage for the mixture. The dependence of our results on $q$ and $m$ is in complete qualitative agreement with 
the results reported in \cite{kn.meakin} ( see Fig. (3)).
 Further, Meakin and Jullien also confirm Feder's law for the small particle species, and also find an
 exponential approach of the coverage by the large ones. In the latter case, an algebraic pre-factor, as
 found here (Eq. (\ref{eq:asym_large})), might be hard to identify in their simulational data. 

In this context it is interesting to verify Palasti's conjecture, stating that the jamming coverage 
$\theta_d$ in d dimensions can be obtained from the one dimensional result, i.e.
\begin{equation}
  \label{eq:palasti}
\theta_d(\infty)\approx (\theta_1(\infty))^d\,.
\end{equation}
Although the strict equality (as conjectured by Palasti \cite{kn.palasti}) is  not valid, it is often found that the 
relation provides a remarkably good estimate {\cite{kn.privman}. Taking the values from figure 3 of the paper by Meakin
 and Jullien for the jamming coverages by a binary mixture of disks, we find a deviation of at most 5\% from 
the estimates provided by our one-dimensional data and Eq. (\ref{eq:palasti}. The virtue of the present
work is the excellent qualitative match of almost every espects of 1d and 2d results. This confirms 
that 1d model indeed is very useful in gaining information about heigher dimensions despite its simplicity.

In conclusion we have shown that an analytical model for the competitive random sequential adsorption (CRSA) of a 
binary mixture in one dimension predicts a jamming limit that is always larger than for the uni-sized case. In the 
limits $q=0$, $q=1$, as well as $m=1$ we recover the classical result $\theta_R=0.748$ for the uni-sized RSA process. 
In all other cases, i.e.\ for true binary mixtures, we find a larger jamming limit in theory and simulation. In general,
 our analytical results are in excellent agreement with direct numerical simulations. The time-asymptotic approach of 
the jamming limit is dominated by the contribution of the small particles, and we confirm the $1/t$ behavior predicted
 by Feder's law in one dimension. The large particles reach their contribution to the jamming limit exponentially with
 an algebraic pre-factor.

\noindent This work was supported by the Deutsche Volkswagen-Stiftung (B.\ B.) and the Humboldt-Foundation (M.\ K. \ H.).

\end{document}